# Analyzing the Dynamic Relationship between Tumor Growth and Angiogenesis in a Two Dimensional Finite Element Model

Eun Bo Shim[1], Yoo Seok Kim[1], and Thomas S. Deisboeck [2, *]

[1] Department of Mechanical and Biomedical Engineering, Kangwon National University, Chuncheon, Kangwon Province, Republic of Korea; [2] Complex Biosystems Modeling Laboratory, Harvard-MIT (HST) Athinoula A. Martinos Center for Biomedical Imaging, Massachusetts General Hospital, Harvard Medical School, Charlestown, MA 02129, USA.

**Running Title:** 2D FEM Tumor Angiogenesis Model

**\*Corresponding Author:**

Thomas S. Deisboeck, M.D.
Complex Biosystems Modeling Laboratory
Harvard-MIT (HST) Athinoula A. Martinos Center for Biomedical Imaging
Massachusetts General Hospital-East, 2301
Bldg. 149, 13th Street
Charlestown, MA 02129
Tel:   617-724-1845
Fax:   617-726-7422
Email: deisboec@helix.mgh.harvard.edu

## ABSTRACT

Employing a novel two-dimensional computational model we have simulated the *feedback* between angiogenesis and tumor growth dynamics. Analyzing vessel formation and elongation towards the concentration gradient of the tumor-derived angiogenetic basic fibroblast growth factor, bFGF, we assumed that prior to the blood vessels reaching the tumor surface, the resulting pattern of tumor growth is symmetric, circular with a common center point. However, after the vessels reach the tumor surface, we assumed that the growth rate of that particular cancer region is accelerated compared to the tumor surface section that lacks neo-vascularization. Therefore, the resulting asymmetric tumor growth pattern is biased towards the site of the nourishing vessels. The simulation results show over time an increase in vessel density, a decrease in vessel branching length, and an increase in *fracticality* of the vascular branching architecture. Interestingly, over time the fractal dimension displayed a *sigmoidal* pattern with a reduced rate increase at earlier and later tumor growth stages due to distinct characteristics in vessel length and density. The finding that, at later stages, higher vascular fracticality resulted in a marked increase of tumor slice volume provides further *in silico* evidence for a functional impact of vascular patterns on cancer growth.

# 1. INTRODUCTION

Despite substantial progress in characterizing the mechanisms that control tumor angiogenesis, following the seminal work by Folkman (1971) and others, the dynamical regional *feedback* between tumor growth and neovascularization is still not fully understood. Since this is due, in large parts, to the complexity of the biological processes involved, *in silico* modeling, which allows for reproducibly altering parameters separately or in combination, can contribute here. Thus, not surprisingly, there have already been a number of theoretical studies on tumor angiogenesis in which continuum or discrete models were used.

In continuum models, only the distribution of endothelial cells is considered while vascular networks are not included. Chaplain *et al.* (1996, 1997) then presented two- and three-dimensional models of tumor angiogenesis using a 'hybrid' by combining both discrete and continuum methods. To determine the movement of the sprouting tips of endothelial cells, the authors solved partial differential equations for the concentration of a so called tumor angiogenesis factor (TAF), of fibronectin and endothelial cell density by using the finite difference method. The formation and growth of vessel sprouts were approximated using a stochastic process that was based on the distribution of both TAF and fibronectin; yet, Chaplain *et al*. also assumed that TAF secretion was constant over time. The numerical solutions of such models can be compared to experimental data, and cellular mechanisms can be incorporated readily into new mathematical models. Alarcon et al. (2003) then developed a mathematical model which showed the influence of blood flow and red blood cell heterogeneity on tumor growth and angiogenesis. Furthermore, Serini et al. (2003) provided a model that included chemoattraction and cell-cell interaction for identification of key parameters in the early stages of the vascular network assembly. Recently, Zheng *et al*. (2005) have computed tumor angiogenesis using a formulation similar to Chaplain, but they modeled the nutrient distribution assuming a spatial variation of TAF within viable and necrotic tumor regions. In this context, we note also that Gazit *et al.* (1995) already employed fractal theory to compute the vessel networks that surround a tumor and computed the hemodynamics within these vessel structures. Grizzi et al. (2005) introduced the surface fractal dimension to explain the geometric complexity of cancerous vascular networks. They showed that the surface fractal dimension significantly depends on the number of vessels and their patterns of distribution. However, these previous studies restricted growth to the discrete mesh points of the computational lattice. Conversely, Tong *et al.* (2001) developed a two-dimensional angiogenesis model in which they assumed a biased random motion of endothelial cells in an effort to examine the transport of angiogenic factors in the rat cornea; vessel growth in their study was independent of the computational mesh unlike in the previous approaches and as such, tumor angiogenesis was implemented in a more realistic and efficient manner. We add that another detailed model of tumor angiogenesis has been proposed by Levine et al. (2001), which included the biochemical processes that involve angiostatin, the biased movement of endothelial cells, and transport diffusion equations of molecular species in porous media. Finally, a specific inclusion of cell metabolism has been made by Scalerandi et al. (2001) in an effort to model the local interaction between cells and the vascular system.

However, one of the main limitations of all these previous works is that the quantity of angiogenic factors that are released from the tumor cells is assumed to be *constant*. This assumption is rather unrealistic, since the increase in tumor volume and thus the quantity of angiogenic factors that are released by it should be linked, i.e. *vary dynamically*. We have therefore introduced a tumor angiogenesis model adopting a pattern of a growing brain tumor (Shim *et al.*, 2005). In this previous study, the angiogenetic factor bFGF was secreted with a time varying pattern and led to symmetric cancer growth. The model's primary shortcoming was that it failed to close the feedback loop that the vessel's emergent architecture should have on regional tumor growth behavior. Therefore, in this study here, we present a computational model of tumor-induced angiogenesis that now includes this crucial *link* between vessel architecture and regional tumor growth dynamics. As before, we have again adopted the model of a virtual brain tumor that was presented in Kansal *et al.* (2000) and employed it for the pre-vascular stage, i.e. prior to any blood vessel reaching the expanding tumor. However, a different growth model is then applied after the first vessel reaches the tumor surface. We chose basic fibroblast growth factor (bFGF) as angiogenesis factor because the concentration of bFGF is reportedly proportional to the increase in malignancy and vascularity of high-grade gliomas (Takahashi *et al.*, 1992). Some parameter values for bFGF-induced angiogenesis were obtained from Tong *et al.* (2001), whereas e.g. the 50 pg/$10^5$ cells per 24 h as bFGF production rate of human U87 glioma cells was taken from Zagzag *et al.* (1990). The finite element method was used to solve the convection-diffusion equation for the concentration of bFGF; this method is a convenient way in dealing with the complex geometry of real biological phenomena. Both vessel formation and sprout elongation were simulated using a stochastic process, much like in the aforementioned studies.

## 2. MODEL

General setup

The two-dimensional setup is depicted in **Figure 1** and consisted of a 171 x 171 rectangular mesh lattice.

**Figure 1**

**Table 1**

In here, $L_{domain}$ and $R_{PV}$ represent the total computational domain and the radius of the parent vessel measured from the center of the tumor, respectively. The computational domain consists of three regions: necrotic and viable tumor regions, and surrounding healthy tissue. For the simulation of tumor growth in the computational domain we used the following assumptions for these regions:

1. In the viable region ($\Omega_V$), bFGF is produced according to the tumor's dynamic growth pattern as reported for the case of a virtual brain tumor by Kansal *et al.* (2000; **Table 2**). Eventually, the secreted bFGF diffuses also into the healthy non-tumorous tissue (and into the tumor's necrotic region). As such, there is bFGF production, diffusion, and time-elapsed decay in this viable region.

2. In the necrotic region ($\Omega_N$), we assume the production of bFGF to be zero; however, as stated above, the bFGF produced in the viable region of the tumor diffuses to the necrotic region and decays over time.

3. In the healthy tissue region ($\Omega_H$), we assume that there is diffusion of bFGF, decay of bFGF due to degradation and uptake by endothelial cells. For the diffusion boundary ($\xi_D$), we also assumed a no flux condition similar to Tong *et al.* (2001).

Transport equations for basic fibroblast growth factor (bFGF)

The transport equation of bFGF within the three regions depends on the characteristics of each domain. A general governing equation for bFGF transport can be derived as follows:

$$\frac{\partial C}{\partial t} = \mu_1 Q_p + \mu_2 D \cdot \left( \frac{\partial^2 C}{\partial x^2} + \frac{\partial^2 C}{\partial y^2} \right) - \mu_3 k \cdot C - \mu_4 u \cdot L \cdot C \qquad (1)$$

where $Q_p$, $C$, $D$, $k$, $u$, and $L$ represent the bFGF production rate, the concentration of bFGF, the diffusion coefficient of bFGF, the rate constant of bFGF degradation, the rate constant of bFGF uptake, and the vessel density (defined as the total vessel length per unit area), respectively. The constants used are listed in **Table 2**. The coefficients $\mu_1$, $\mu_2$, $\mu_3$, and $\mu_4$ represent on-off style identifiers (set as 0 or 1) and vary according to regional characteristic for the bFGF transport as described in **Table 3**. $Q_p$ is non-zero in the viable tumor region whereas bFGF uptake by endothelial cells is zero in this region. The transient value of $Q_p$ is given in the last row of **Table 1** based on the experimental observation that human U87 glioma cells produce 50 pg of bFGF (per $10^5$ cells) over 24 h (Zagzag *et al.*, 1990) and the time-dependent viable tumor cell number listed in **Table 1**.

The bFGF diffusion coefficient D was taken from Tong *et al.* (2001) and assumed to be constant over all regions.

**Table 2**

**Table 3**

For the computational domain, application of the Galerkin finite element discretization for Eq. (1) yielded the matrix equation with the coefficient matrix, the vector of unknown nodal variables of C and the external driving forces. This matrix equation was solved using an incomplete conjugate gradient method (Kershaw, 1978).

Sprout formation and elongation

Within the surrounding normal tissue, the initial response of the endothelial cells to bFGF is *chemotactic*, i.e. migration along the bFGF-concentration gradient and thus towards the angiogentic factor-releasing tumor. This process leads to the formation of capillary sprouts that continue to grow in length towards the growing tumor, guided by the motion of the leading endothelial cell at the tip of the sprout. Much like Tong *et al.* (2001), we also introduced a threshold function f(C) to account for the effect of the bFGF concentration on vessel sprout formation and elongation as follows: the values of the function are zero below the threshold concentration $C_t$ and increase exponentially with a limiting value of 1. The value of $C_t$ is also represented in **Table 2**. To approximate sprout formation, we assume that it is a stochastic process with biased angiogenesis pattern towards the tumor. The probability $\bar{n}$ for the formation of one sprout from a vessel segment in a time interval between t and t+Δt is proportional to Δt, the segment length Δl, and the threshold function. Here, the proportional constant, $S_{max}$, denotes a rate constant that determines the maximum probability of sprout formation per unit time and vessel length. The growth of a sprout is determined by the locomotion of its tip, while the geometry of a sprout depends on the tip trajectory (Tong *et al.*, 2001). Specifically, the direction of sprout growth at each time step depends on two unit vectors: the direction of growth in the previous time step and the direction of the concentration gradient of the angiogenic factors. This is due to the fact that sprout growth depends on endothelial cell migration, which has a tendency to persist in the same direction as in the previous time step. To reflect the effect of extracellular matrix on (haptotacic) cell migration, we assume that the angle of deviation, $\theta$, is between $\pi/2$ and $-\pi/2$ and that $\tan\theta$ follows a Gaussian distribution with a mean of zero and a variance of $\sigma$. A detailed description of sprout formation and elongation can be found in Tong *et al.* (2001). The constants represented in the equations are described in **Table 2**.

Brain tumor growth model

We employ a previously developed brain tumor model that has four distinct growth stages within a virtual patient, namely multicellular spheroid, 1$^{st}$ detectable lesion, diagnosis, and death (Kansal *et al.*, 2000). To approximate data in each of these growth stages, we used the following well known Gompertz equation:

$$V = V_0 \exp\left(\frac{A}{B}\left(1 - \exp(-Bt)\right)\right) \qquad (2)$$

where A and B are parameters, and $V_0$ is the initial tumor volume. The quantity of bFGF release at each of the two 'early' growth stages is summarized in **Table 1**. Assuming that tumor growth would be *spherical* during the *avascular* state but would then develop a *biased* shape towards the parent vessel during the *vascular* state, we then investigated a 1.0-mm-thick, circular slice of the tumor for our 2D model. As a first approximation, we have calculated the amount of bFGF at each of the two stages for the total amount of viable tumor cells (i.e., proliferative and quiescent) using the cell numbers reported by Kansal *et al.* (2000)[1]. Further, as soon as the first vessel reaches the tumor surface, we assume that the part of the tumor adjacent to this vessel obeys vascularized tumor growth dynamics. While this particular tumor region will therefore display a rather rapid radius increase, the rest of the tumor, non-vascularized, continues to follow the slower avascular dynamics of the spheroid stage. This implements the concept that tumor growth is largely symmetrical before the first vessel branch makes contact with the tumor surface; thereafter, growth is arguably more asymmetrical for that particular tumor region. Reflecting early experimental work by Folkman (1971), the 'first' vessel branch was set to reach the tumor surface when the radius of the tumor is 1.0 mm; afterwards, the tumor begins to grow eccentrically since biased towards the nourishing blood vessel. For computation of the asymmetric growth, we assumed that there are two regions as shown in **Figure 2**: vascularized and non-vascularized regions.

---

[1] This model focused on describing the tumor's solid core, hence included proliferative and non-proliferative alive as well as dead tumor cells, yet did not simulate migrating tumor cells explicitly.

**Figure 2**

**Figure 3**

Note that there are several regional tumor centers developing in the vascularized region, from which tumor growth is augmented; in this paper we however keep the rate at which the radius increases, which is a variable computed from the time-varying slice volume data, uniform for all these regional tumor centers (follow-up work will relax this assumption to better reflect the emergence of tumor heterogeneity). **Figure 3** represents the overall tumor radius computed from the Gompertz equation (Eq. 2) with the parameters adjusted to match the data listed in **Table 1**. Here, the tumor slice volume is computed from this radius with a constant slice thickness of 0.1 mm. The main assumption of the asymmetric tumor growth in the present study is that the volume growth difference between vascular and avascular state contributes to the asymmetric growth of the tumor after the $1^{st}$ vessel reaches the tumor surface. The following paragraphs briefly describe the schematic behind the asymmetric growth procedure during the vascular state per each time step:

1. When the $1^{st}$ blood vessel reaches the tumor surface, the sites that are 'touched' by these vascular branches will be specified as *local* tumor centers. In **Figure 2(a)**, $C_0$ denotes the initial tumor center that is the origin of symmetric tumor growth whereas the others ($C_1$, $C_2$, ..., $C_n$) comprise these local tumor centers. That is, the non-vascularized region maintains the initial tumor center, $C_0$, whereas the vascularized regions continue with local tumor centers, i.e., $C_1$, $C_2$, ..., $C_n$.

2. After elapse of one time step, the tumor boundary expanded into the dotted line in **Figure 2(a)**. Here, *t* stands for time and its superscript denotes time index. In this first approximation, we assume that the local tumor centers have the same radius increase ($\delta$) whereas the non-vascularized region distinguishes itself through a radius increase of $\delta_0$. This radius increase of $\delta_0$ is obtained from curve A in **Figure 2(b)**. The radius increase $\delta$ in the vascularized region (curve B, **Figure 2(b)**) was obtained to satisfy the area of the vascularized tumor region equal to the computed tumor slice volume of the Gompertz equation curve in **Figure 3**.

3. For the vascularized tumor area it then follows that it sustains augmented volumetric growth and thus harbors more viable cancer cells.

4. Consequently, this gain in viable tumor cells leads to an increase in bFGF production in the very same vascularized tumor area.

5. Lastly, this increased bFGF secretion recruits even more blood vessels towards an already vascularized area and thus 'closes' the *feedback loop* back to (3.).

Branching pattern analysis

To quantify the vessel branching patterns within the computational domain, we calculated the number of branching points within a given region of interest (ROI, see **Figure 1**). A ROI represents a rectangular region with 2.5mm × 2.5mm and a branching point is defined as the site at which one vessel splits into two new branches. For each ROI, the total number of branching points within the ROI was calculated. To account for temporal variation of vessel growth, we considered the average branching length, number of branching points and vessel density in the ROI. Also the fractal dimension analysis for the vessel structure is analyzed to track the dynamic change in vessel formation. In brief, the fractal dimension in this study is based on the box counting method that is derived from the theoretical works by Baish and Jain (2000). A total of 20 different boxes with a specific width of the total computational domain are used. We then computed the corresponding box numbers to cover all vessels in the domain and this was plotted according to the box numbers. The linear slope of the logarithmic graph represents the fractal dimension.

# 3. RESULTS

We have implemented the algorithm using Fortran. One run took approximately 28 hrs using a Pentium 586 PC with 2.33 GHz clock speed. To investigate the effect of the location of the initial vessel formation on the emerging vascular network, we have run 10 cases with a random choice in the initial sprout location. The circumferential angles of these locations are listed in the **Table 4**. However, for brevity we focus in the following section on the case with $\theta_1 = -103°$, $\theta_2 = -135°$, $\theta_3 = -168°$ (except for **Figure 10**, as stated). In the following, we describe the results in detail.

**Table 4**

To begin with, the tumor-secreted bFGF concentration distribution (**Figure 4**) displayed a radial isotropic gradient at t = 1,656 hrs and 2,200 hrs (**Figs. 4(a)-(b)**) when the vessels were still rather short. However, after the vessel architecture has expanded and moved closer to the tumor surface, the overall endothelial cells' bFGF consumption increased markedly and, consequently, the radial gradient of the concentration distribution of bFGF ceased to diffuse isotropically. Specifically, the concentration of bFGF decreased on the tumor side adjacent to the blood vessels while it remained relatively high on the opposite side (**Figs. 4(c)-(e)**). The plots also show that due to its ongoing production the maximum value of the bFGF concentration increases prior to the first vessel branch reaching the tumor surface (top value in color bar, **Figs. 4(a)-(b)**). While the aforementioned endothelial consumption leads to a temporary reduction of the maximum bFGF value (**Figs. 4(c)-(d)**), overall, the bFGF concentration in the tumor slice volume increases again later on (**Figs. 4(e)-(f)**).

**Figure 4**

**Figure 5**

It is noteworthy that, although originating at the same time point, the middle branch showed relatively less branching due to 'competition' with the others for limited amounts of released bFGF. In our simulation, the 1$^{st}$ vessel branch reaches the tumor surface at t = 2,590 hrs (**Figure 4(c);** t = 2,800 h) while at t = 3,800 hrs, numerous vessel branches close to the tumor surface yield a so-called "brush-border effect" with higher vessel densities within that tumor region (**Figure 4(f)**).

Structural characteristics of tumor angiogenesis are then depicted in **Figure 5**. First, we note that the branching length diminishes exponentially over time (**Figure 5(a)**) while the number of branching points and the overall vessel density show an exponential increase (**Figs. (b)-(c)**). The results therefore confirm that the number of vessel branches increased dramatically as soon as the first vessels reached the tumor surface (t > 2,590 hrs), enabling the transition to the vascular growth stage.

**Figure 6**

**Figure 7**

**Figure 8**

Analysis of the vessel architecture's fractal dimension is represented in **Figure 6**, which illustrates a gradual increase over time. Specifically, during the avascular state (**Figs. 6(a)-(b)**) the fractal dimension is less than 1. However, as more vessels reach the tumor surface, the dimension increases beyond 1 and reaches 1.4 at t = 3,800 hrs. While **Figure 7** summarizes this change of fractal dimension over time, **Figure 8** displays the variation of the fractal dimension versus the tumor slice volume. The latter result demonstrates the nourishing impact vascularization has when, at a fractal dimension of 1.2, the tumor's growth rate is dramatically augmented. Interestingly, the vessel architecture's fractal dimension appears to approach an asymptotic value at larger tumor slice volumes. Taken together with **Figure 5**, this indicates that at later stages, the emergent vascular infrastructure seems to rely increasingly on relatively short and straight vessels.

**Figure 9**

**Figure 10**

To assess the effect of the variance in the sprout's deviation angle (**Table 2**) on angiogenesis, and thus to evaluate the robustness of the results, we simulated two more cases for the variance $\sigma = 0.3$ (**Figure 9(a)**) and $\sigma = 0.7$ (**Figure 9(c)**), respectively, and compared it with the standard case of $\sigma = 0.5$ (**Figure 9(b)**). While, according to the increase of variance in the deviation angle, the circumference length occupied by the vascularized tumor region indeed increased, we found no significant change in the fractal dimension (**Figure 10**), indicating that the latter is invariant to the deviation angle.

**Figure 11**

Finally, to investigate the dependency of the fractal dimension on the tumor radius at the time the $1^{st}$ vessel docks, we simulated three more cases with varying tumor radii (**Figure 11**). Similar to the standard case depicted in **Figure 7**, the fractal dimension increased again yet was saturated at later times. While the lag phase seems to last longer (**Figure 11(c)**) for larger tumor radii, the curve's overall *sigmoidal* pattern is robust.

## 4. DISCUSSION & CONCLUSIONS

A more detailed understanding of tumor angiogenesis is of paramount interest for clinical cancer research in an effort to develop more effective anticancer therapies (for a review see e.g. Carmeliet & Jain, 2000). We firmly believe that *in silico* research can help guide experimental works and so, in here, we have presented a new computational method to simulate tumor angiogenesis in two dimensions. As an example for angiogeneic factors, we used basic fibroblast growth factor or bFGF. Its expression reportedly correlates with the degree of malignancy and vascularity in gliomas (Takahashi *et al.*, 1992). For the analysis of the spatio-temporal distribution of bFGF, its conservation equation was solved using the finite element method. Unlike in previous computational and mathematical studies, here, we have taken into account the *feedback* between vascular supply and tumor growth. Specifically, employing data from a previous study that describes the virtual growth of a malignant brain tumor over several scales of interest (Kansal *et al.*, 2000) we have focused on monitoring the dynamics of the tumor's bFGF production and its effect on the emergent vessel patterns.

The simulation results confirm that a tumor-secreted angiogeneic factor influences the patterns of vascular architecture and that such dynamic neovascularization can impact tumor growth patterns (**Figures 4, 8**), thus closing the assumed feedback loop. While these results are somewhat expected given the setup of the underlying algorithm, to our knowledge, an *in silico* model that simulates these relationships properly has not been developed yet. Aside from this technical advancement, the fact that the results show a clear shift in vessel *structure*, i.e. fractality, at the transition to the vascular tumor growth stages deserves a more detailed discussion. For instance, applying West *et al.*'s Universal scaling law (2001) to tumors, Guiot *et al.* (2003) have previously argued for a dynamic behavior of the so called scaling exponent '*p*' (Guiot *et al.*, 2005). Specifically, that work conjectured that scaling exponent values which exceed ¾ (and are thought to be conveyed by angiogenesis) can be explained with surface-diffusion supplementing neovascularization as prevailing tumor nourishing supply mechanism. These theoretical considerations are well matched now with the result from our computation here, where vessel branching indeed starts at about t = 2,200 hrs (**Figure 4(b)**) with a fractal dimension of about 0.71 (**Figures 6(b), 7**) before it increases well over 1. Taken together, we argue that the scaling exponent *p* at which angiogenesis starts is indeed between 2/3 and 3/4. Interestingly, we also find that over time the fractal dimension shows a rather robust *sigmoidal* pattern with a saturated curve at the earlier and later stages (**Figures 10 and 11**). That is, at the initial stage the fractal dimension slowly increases because the few generated vessel branches are comparably long and straight. This period is followed by a phase of sharp increase in fracticality, prior to displaying a more saturated pattern at the later, established vascular stage which operates with a large number of rather short, straight and space-filling vessels. This not only hints at the fact that vessel fracticality is not limitless but also suggests that – at least temporarily, during the angiogenetic switch – a *more* heterogeneous neovascular architecture may *increase* its functional efficiency to nourish: During that period, a lesser increase in fracticality, by concomitantly utilizing a larger number of shorter and non-fractal microvessels, yields a *more* sustained increase in tumor slice volume (**Figure 8**), indicated also by the larger bFGF concentration this gain in viable tumor tissue can generate (**Figure 4(f)**). It will be intriguing to assess the robustness of this finding after 'functionality' has been introduced explicitly in form of accounting first for a (variety in) vessel diameter (Wesseling *et al.*, 1998) to then model (heterogeneity in) tumor blood flow and blood volume conveyed by the discrete vascular architecture. One goal here would be to *in silico* simulate and analyze the at least for glioblastoma reported (Parikh *et al.*, 2004) correlation between relative cerebral blood volume and abnormal vessel turtuosity (Bullitt *et al.*, 2004). Given the significant effort that is currently underway clinically to image angiogenesis (for a recent review see e.g. Miller *et al.*, 2005), any such interdisciplinary efforts may have substantial applicability.

Admittedly, as a first approximation the model has to rely on multiple simplifications, and future work will need to add complexity on the molecular level both on the tumor and endothelial cells side to enable incorporation of many additionally relevant biomedical data. Secondly, enriching the 'available' cancer cell phenotypes by considering cell motility explicitly will be an important step, particularly also for modeling the spatio-temporal expansion of primary brain tumors more realistically. Lastly, implementing a more specific treatment of the biomechanical properties of tumor and surrounding tissue (see e.g. Boucher *et al.*, 1996) as well as moving the model into three-dimensions represent other avenues that should be pursued. Nonetheless, we argue that if properly expanded this *in silico* platform will offer an exciting new tool to integrative tumor biology research in that it allows for rapid development and refinement of experimentally testable hypotheses related to cancer angiogenesis.

## ACKNOWLEDGEMENTS

This work has been supported by NIH Grants CA113004 and CA085139 from the National Cancer Institute and by the Harvard-MIT (HST) Athinoula A. Martinos Center for Biomedical Imaging and the Department of Radiology at Massachusetts General Hospital.

## FIGURE CAPTIONS

**Figure 1.** Two-dimensional model geometry. Here, $L_{domain}$ is the overall computational length. Depicted are also the location of the parent vessel inside of the domain and the diffusion boundary. $R_{PV}$ represents the radius of the parent vessel location from the tumor center. This figure also illustrates the region of interest (ROI) that is utilized in the analysis (see **Figure 5 (c)**).

**Figure 2.** Schematic of the computational method to calculate the asymmetric tumor growth behavior. Here, the vascularized region displays a relatively faster growth rate than the non-vascularized parts of the tumor. This growth rate is determined by the condition to fit the slice volume of previously reported data (Kansal *et al.*, 2000). (**a**) Asymmetric growth pattern comprise vascularized and non-vascularized tumor regions; (**b**) Schematic of the slice volume growth rate for the case without (curve A) and with angiogenesis (curve B), for the two different growth states, avascular and vascular (compare with "spheroid" and "1$^{st}$ detectable lesion" stages described in **Table 1** (values from Kansal *et al.*, (2000)).

**Figure 3.** Tumor slice volume (left *y-axis*) and corresponding relative (*) concentration of basic fibroblast growth factor (bFGF; right *y-axis*) over time (*x-axis*). (* = relative to the bFGF concentration at the time the first vessel docks on the tumor).

**Figure 4.** Color-coded contours of the bFGF concentration gradient and resulting discrete vessel structure, with tumor contour in the center, at (**a**) t = 1,656 hrs (= "Spheroid" stage, **Table 1**), (**b**) t = 2,200 hrs, (**c**) t = 2,800 hrs (~ "1$^{st}$ Detectable Lesion" stage, **Table 1**), (**d**) t = 3,000 hrs, (**e**) t = 3,400 hrs, and (**f**) t = 3,800 hrs.

**Figure 5.** Variations of structural properties of the vascular architecture over time (*x-axes*): (**a**) branching length (average), (**b**) number of branching points for 24 hrs, and (**c**) vessel density of ROI (compare with **Figure 1**).

**Figure 6.** Fractal dimension of the vascular architecture during tumor growth, in a single run, at (**a**) t = 1,656 hrs, (**b**) t = 2,200 hrs, (**c**) t = 2,800 hrs, (**d**) t = 3,000 hrs, (**e**) t = 3,400 hrs, and (**f**) t = 3,800 hrs.

**Figure 7.** Fractal dimension of the vascular architecture (*y-axis*) over time (*x-axis*). The error bars represent 10 runs with random vessel seed.

**Figure 8.** Fractal dimension of the vascular architecture (*y-axis*) versus tumor slice volume (*x-axis*). The error bars represent 10 runs with random vessel seed.

**Figure 9.** Color-coded contours of the bFGF concentration gradient and discrete vessel structure at t = 3,800 hrs, with variance in the deviation angle of the vessel sprout: (**a**) $\sigma$ = 0.3, (**b**) $\sigma$ = 0.5 (i.e., the standard case, as detailed in **Table 2**), and (**c**) $\sigma$ = 0.7.

**Figure 10.** Fractal dimension of the vascular architecture (*y-axis*) versus time (*x-axis*) with different variance in the sprout's deviation angle (see also **Figure 9**).

**Figure 11.** Fractal dimension of the vascular architecture (*y-axis*) over time (*x-axis*) with varying tumor radii, r, at which the 1st vessel touches the tumor surface: (**a**) r=1.25, (**b**) r=1.5, and (**c**) r=1.75. The error bars represent 10 runs with random vessel seed:

**Table 1.** Approximate concentration of basic fibroblast growth factor (bFGF) released during each of the two consecutive growth stages (derived from a previously published model that simulates the growth of a brain tumor over several orders of magnitude (Kansal *et al.*, 2000). (Note that in this model, "1st detectable lesion" was based on the reported conventional imaging detection limit at the time).

**Table 2.** Values of model constants.

**Table 3.** Values of the on-off style identifiers for the three regions of the tumor as described in the text and referred to in **Figure 1**.

**Table 4.** The circumferential angles (in degree) of the locations of the initial vascular branch that sprouts from the parent vessel.

# FIGURES & TABLES

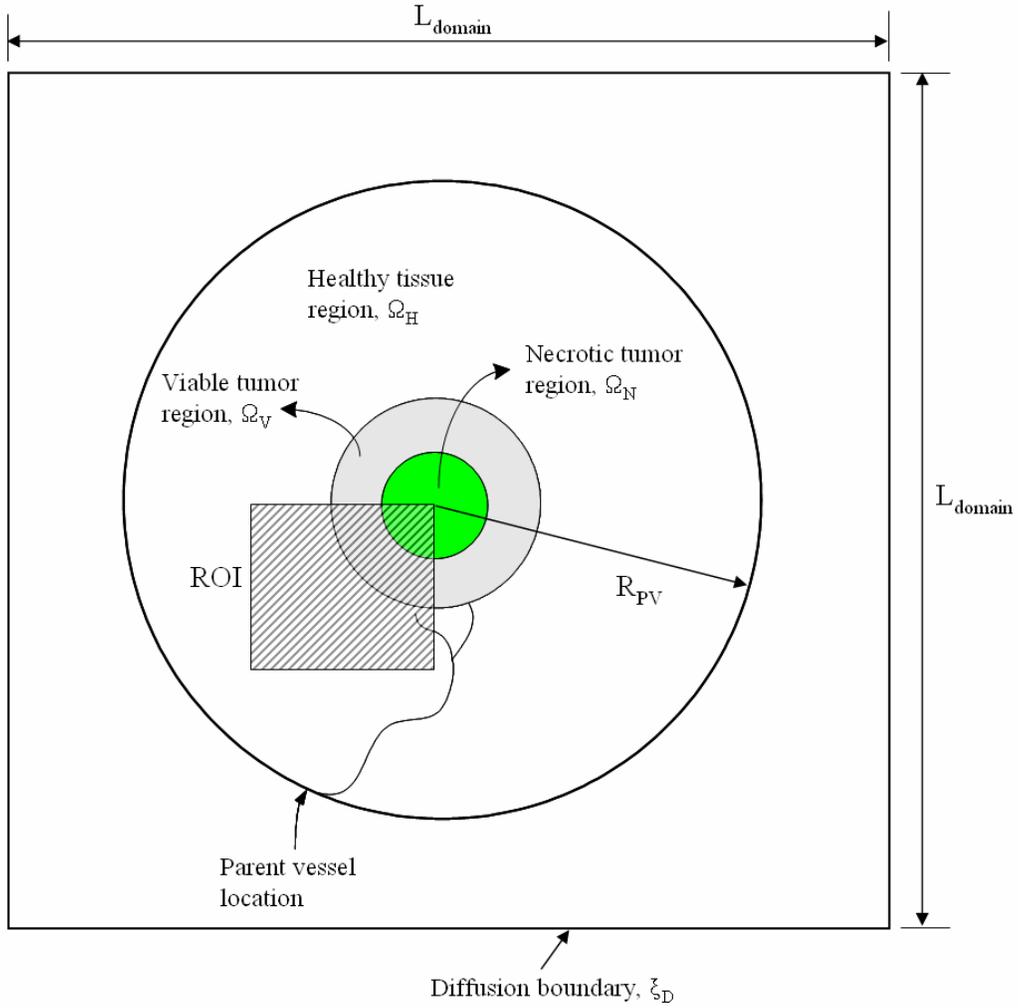

**Figure 1.**

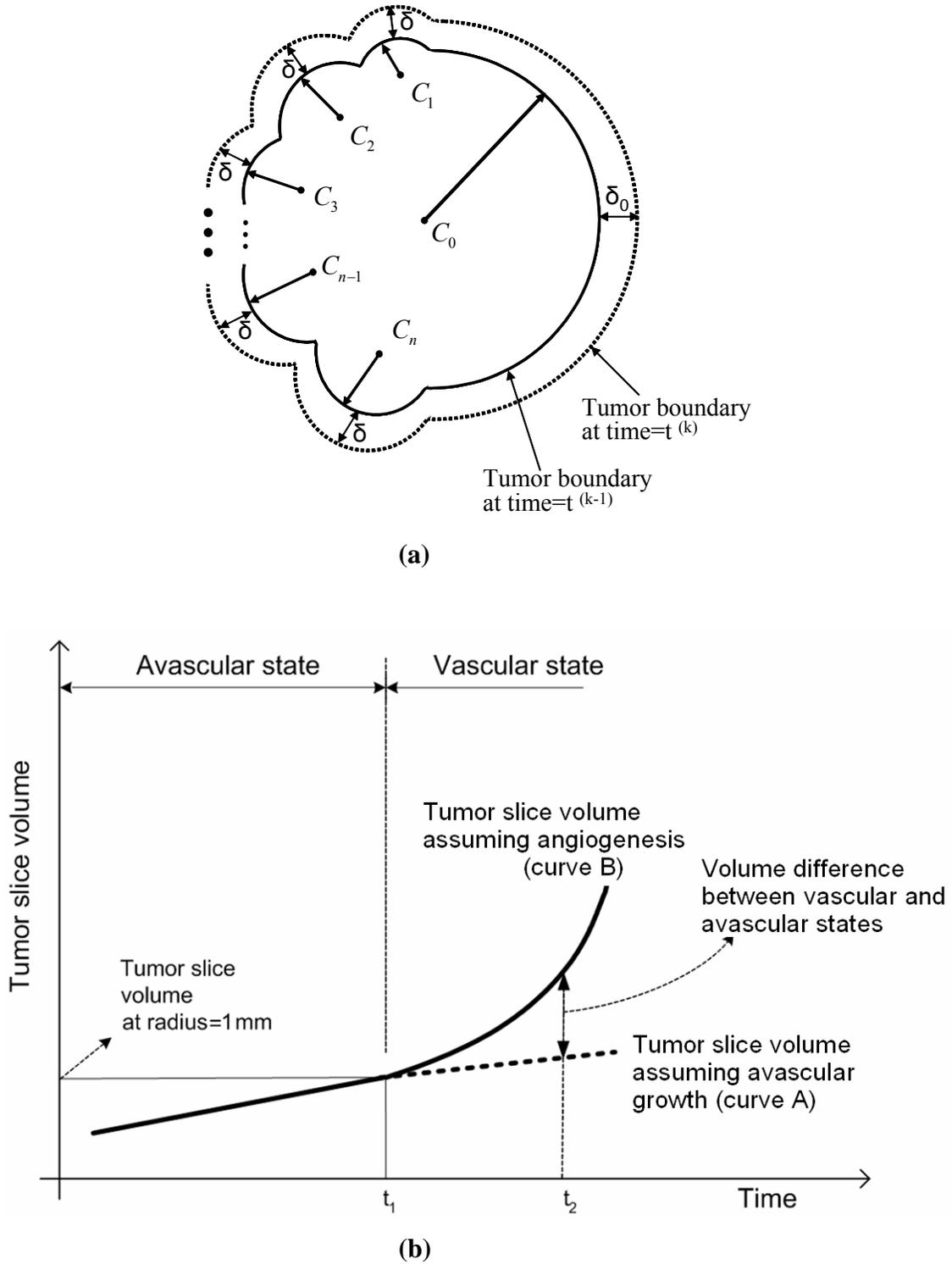

**(a)**

**(b)**

**Figure 2.**

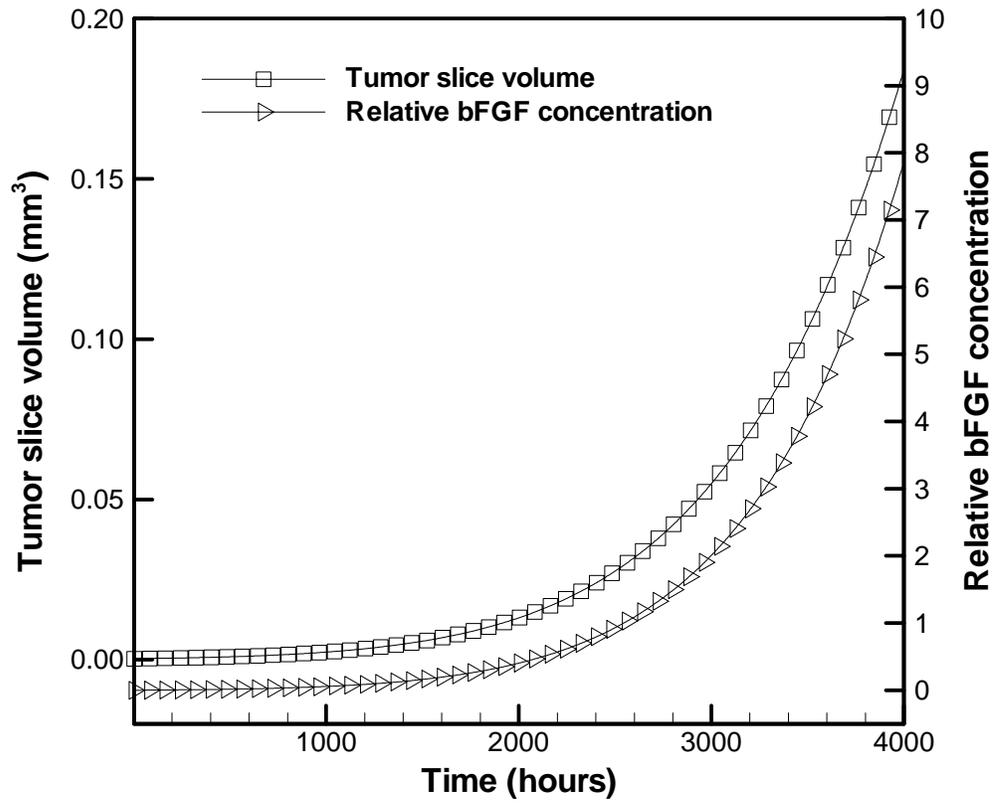

**Figure 3.**

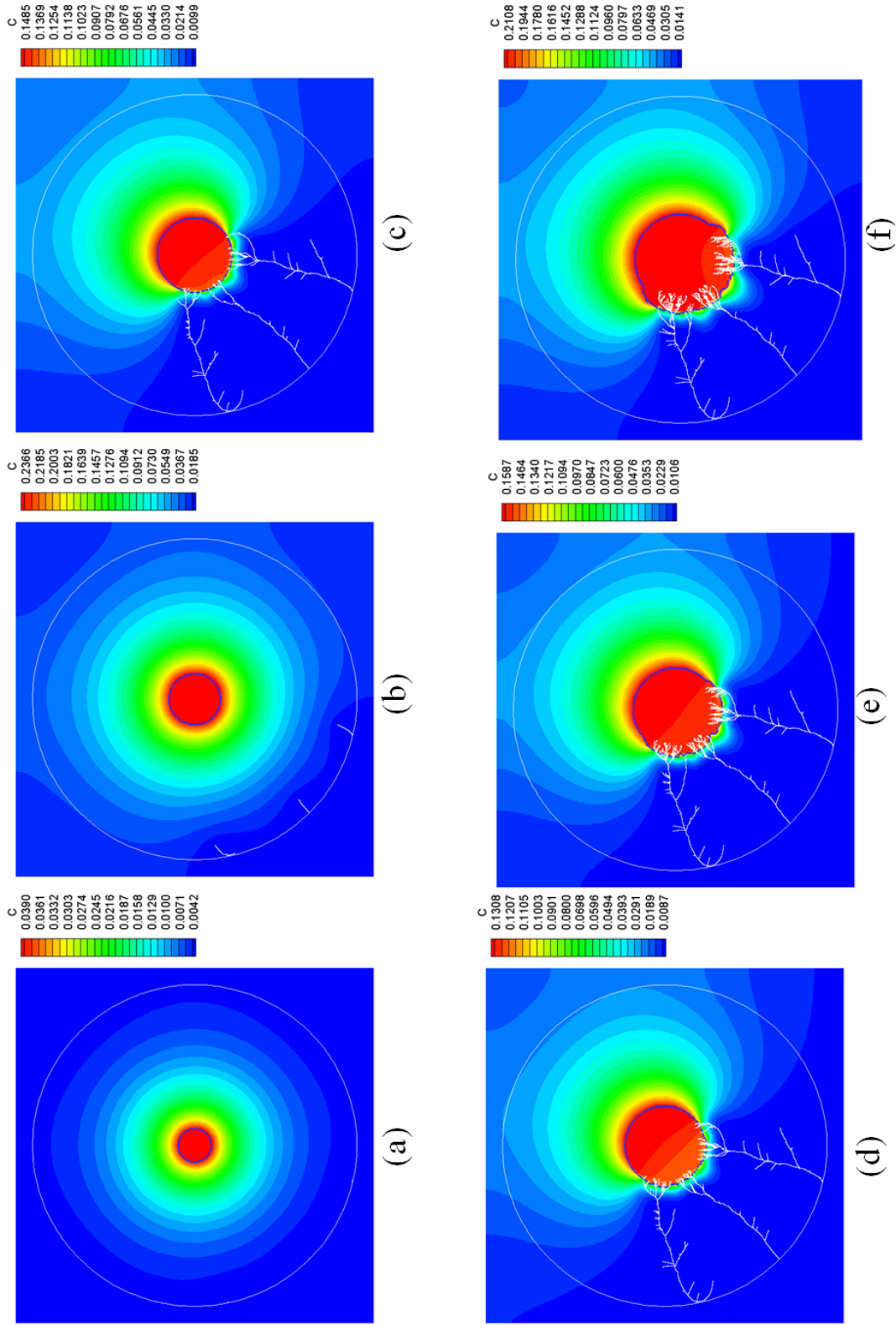

**Figure 4.**

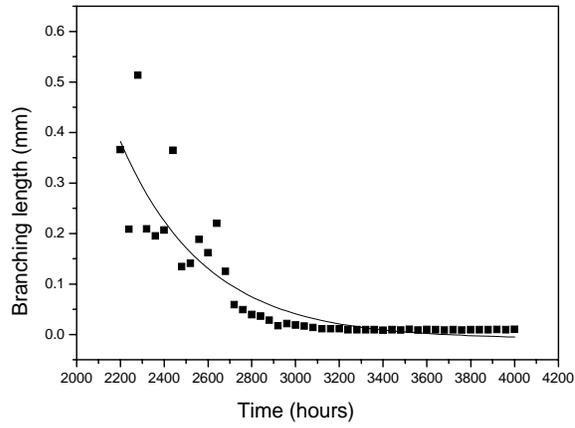

(a)

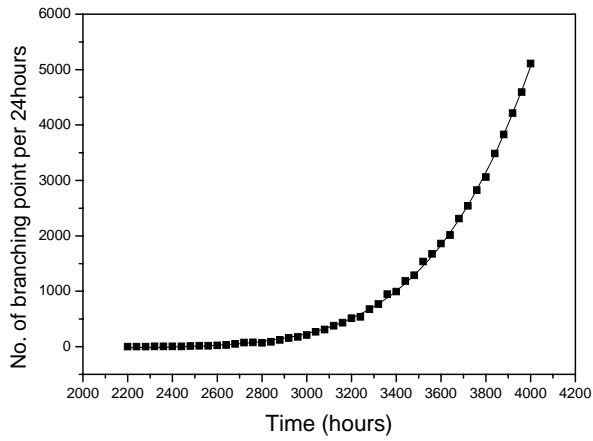

(b)

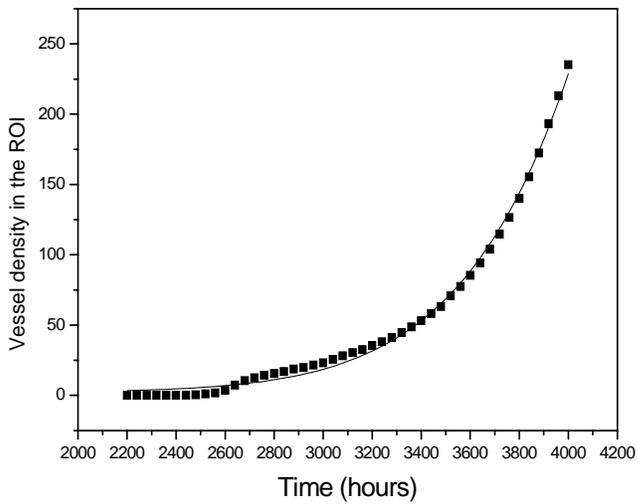

(c)

**Figure 5.**

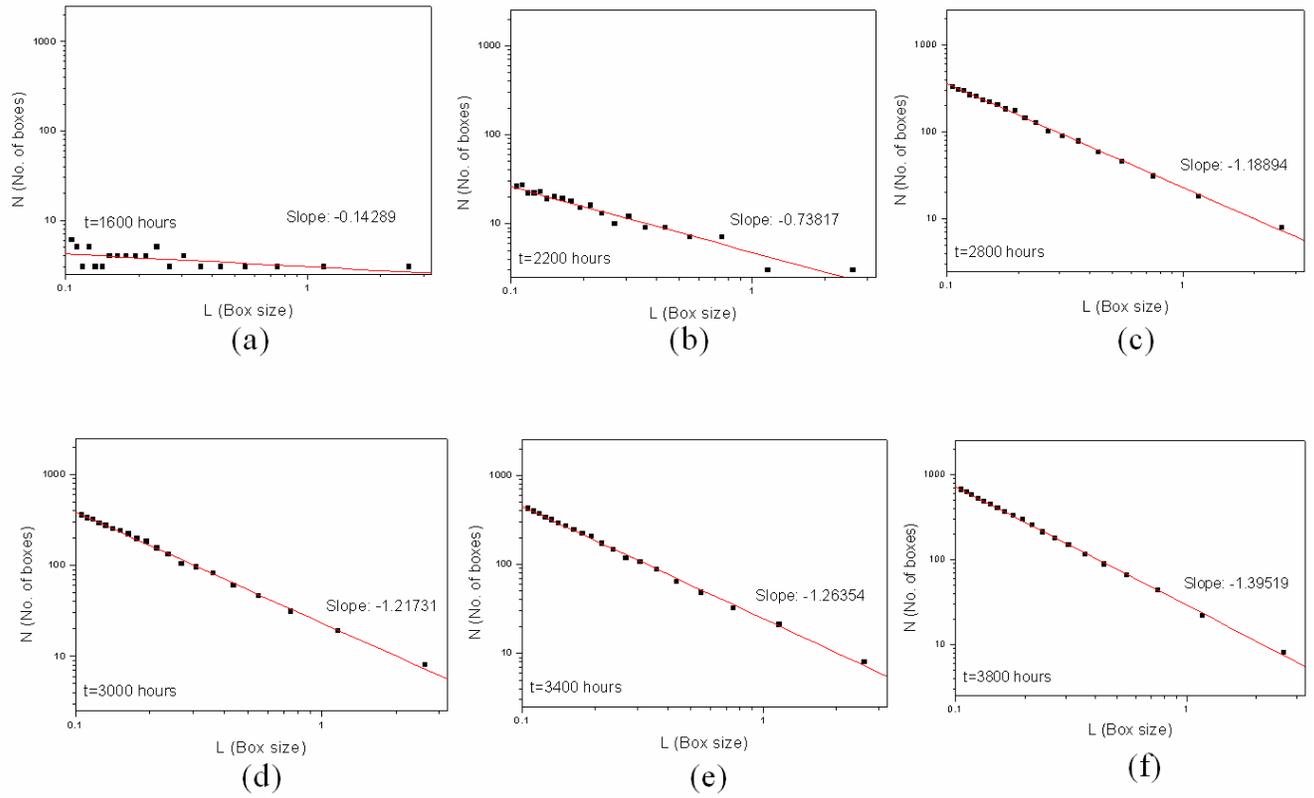

**Figure 6.**

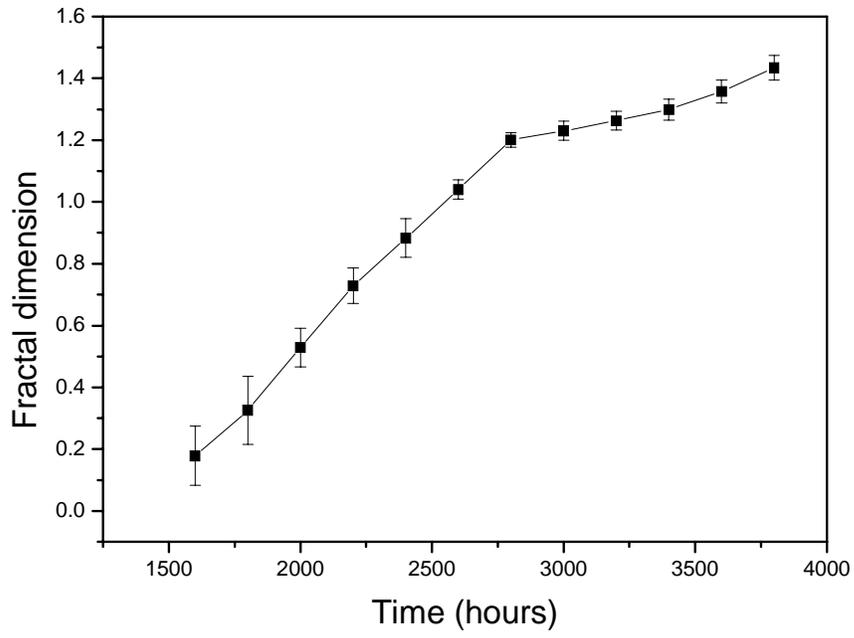

**Figure 7.**

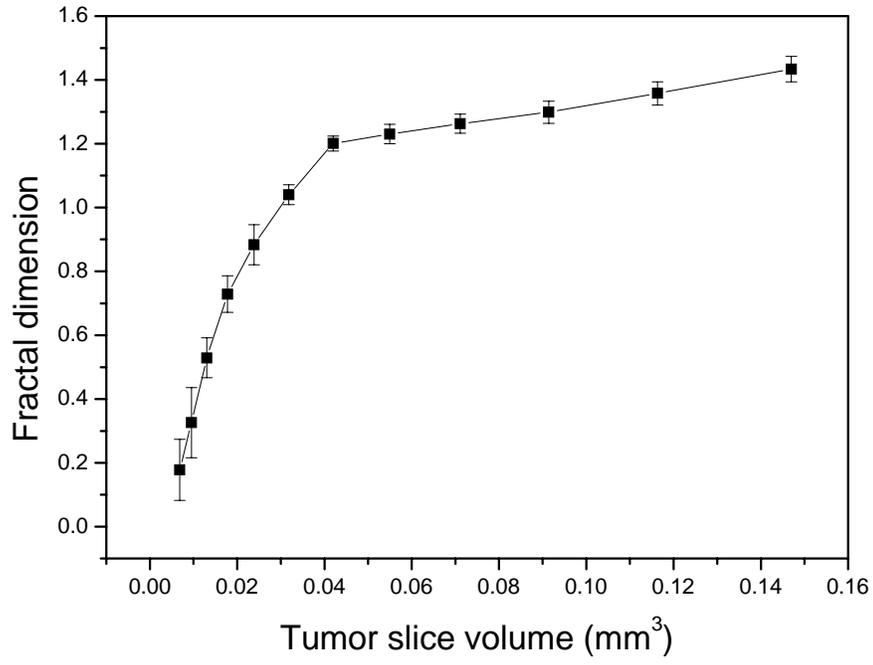

**Figure 8.**

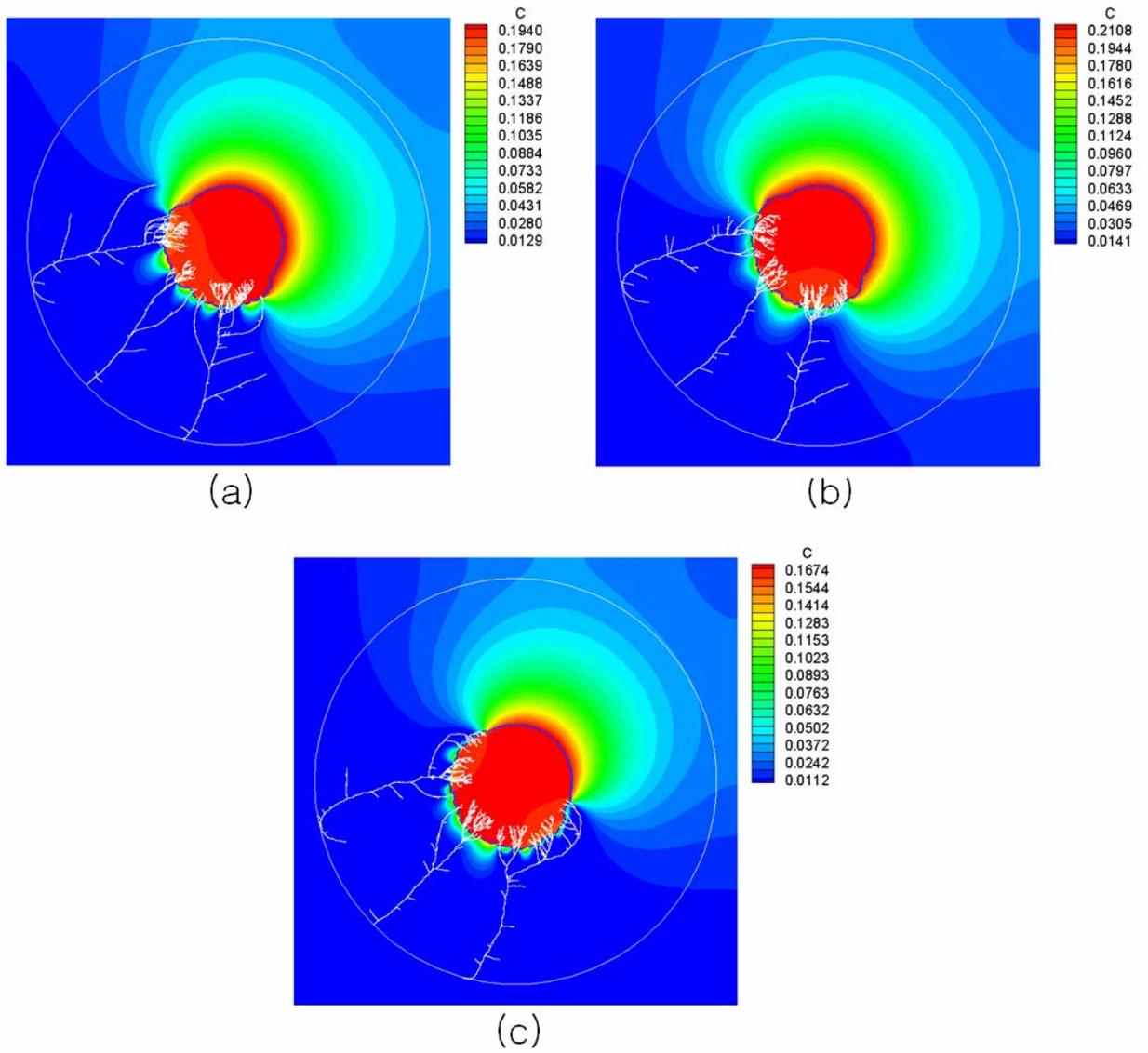

**Figure 9.**

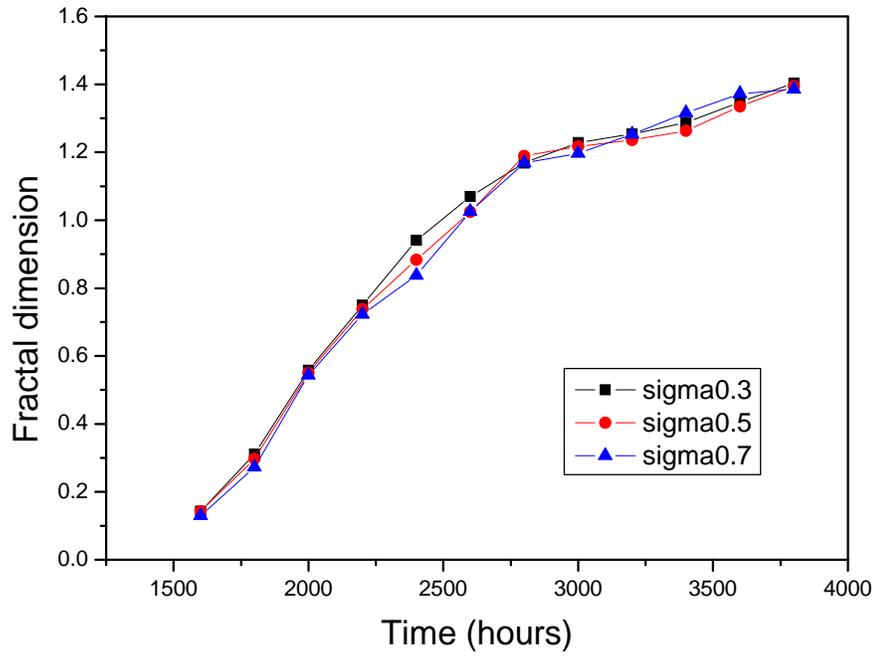

**Figure 10.**

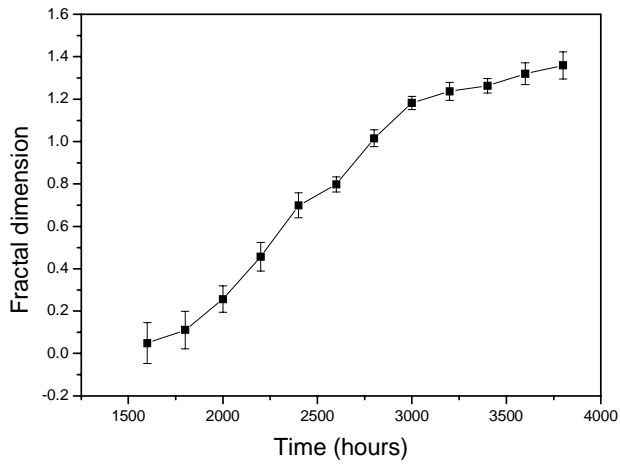

**(a)**

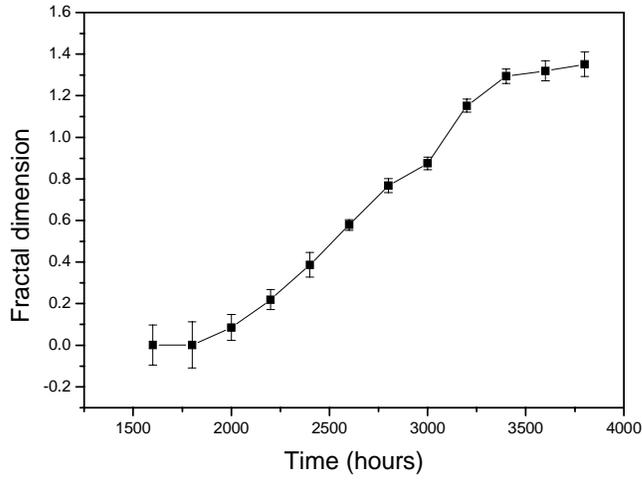

**(b)**

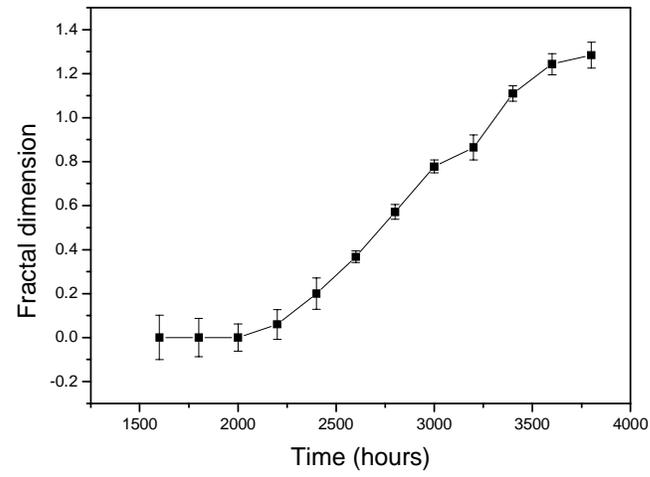

**(c)**

**Figure 11.**

| | Growth Stage | |
|---|---|---|
| | **Spheroid** | **1ˢᵗ Detectable lesion** |
| **Tumor volume data** | | |
|     Total volume ($V$) | $0.5236$ mm$^3$ | $523.6$ mm$^3$ |
|     Total number of cells ($N_{total}$) | $10^6$ | $10^9$ |
|     Radius ($R$) | $0.5$ mm | $5$ mm |
| **Tumor slice volume data** | | |
|     Slice volume ($V_s = \pi R^2 t$) | $0.07854$ mm$^3$ | $7.854$ mm$^3$ |
|     $t$ = slice thickness (= 0.1 mm ) | | |
|     No. of cells per slice volume ($N_{slice}$) | $1.5 \times 10^5$ <br> ($10^6 \times 0.07854/0.5236$) | $1.5 \times 10^7$ <br> ($10^9 \times 7.854/523.6$) |
|     Proliferative *and* quiescent cell volume per slice (cell fraction) | $0.0424$ mm$^3$ <br> (54 %) | $4.0$ mm$^3$ <br> (51 %) |
|     No. of cells per slice volume ($N_{slice}$) | $1.5 \times 10^5$ <br> ($10^6 \times 0.07854/0.5236$) | $1.5 \times 10^7$ <br> ($10^9 \times 7.854/523.6$) |
|     No. of viable cells in slice volume, $N_{viable}$ = (proliferative + quiescent cell fraction) × N$_{slice}$ | $8.1 \times 10^4$ | $7.65 \times 10^6$ |
| Elapsed time of tumor growth (simulation results (Kansal *et al.*, 2000)) | 69 days | 223 days |
| bFGF production rate from $N_{viable}$ tumor cells | 40.5 pg/24 h <br> ($8.1 \times 10^4 \times 50/10^5$) | 3,825 pg/24 h <br> ($7.65 \times 10^6 \times 50/10^5$) |

**Table 1.**

| Constant | Notation | Value | Reference |
|---|---|---|---|
| Diffusion coefficient for bFGF | $D$ | $0.5 \times 10^{-6}\ cm^2\ s^{-1}$ | Tong *et al.* (2001) |
| Rate constant of bFGF uptake | $U$ | $2000.0\ \mu m\ h^{-1}$ | Tong *et al.* (2001) |
| Threshold concentration of bFGF | $C_t$ | 0.001 | Tong *et al.* (2001) |
| Variance of deviation angle for vessel sprout | $\sigma$ | 0.5 | Tong *et al.* (2001) |
| Rate constant of sprout formation | $S_{max}$ | $5 \times 10^{-4}\ \mu m^{-1}\ h^{-1}$ | Tong *et al.* (2001) |
| Rate constant of bFGF degradation | $K$ | $2.89 \times 10^{-2}\ h^{-1}$ | Tong *et al.* (2001) |
| Initial volume in the Gompertz equation (Eq. (2)) | $V_0$ | 0.0042 | Model fit to obtain the previous data (Kansal *et al.*, 2000) |
| Coefficients in the Gompertz equation (Eq. (2)) | A<br>B | 0.0033<br>0.00017 | Model fit to obtain the previous data (Kansal *et al.*, 2000) |

**Table 2.**

| Constant | Values for each region |
|---|---|
| $\mu_1$ | 0  for necrotic region, $\Omega_N$ <br><br> 1  for viable tumor region, $\Omega_V$ <br><br> 0  for healthy tissue region, $\Omega_H$ |
| $\mu_2$ | 1  for necrotic region, $\Omega_N$ <br><br> 1  for viable tumor region, $\Omega_V$ <br><br> 1  for healthy tissue region, $\Omega_H$ |
| $\mu_3$ | 1  for necrotic region, $\Omega_N$ <br><br> 1  for viable tumor region, $\Omega_V$ <br><br> 1  for healthy tissue region, $\Omega_H$ |
| $\mu_4$ | 0  for necrotic region, $\Omega_N$ <br><br> 0  for viable tumor region, $\Omega_V$ <br><br> 1  for healthy tissue region, $\Omega_H$ |

**Table 3.**

| | $\theta_1$ | $\theta_2$ | $\theta_3$ |
|---|---|---|---|
| Case1 | −103.0 | −135.0 | −168.0 |
| Case2 | −92.0 | −120.0 | −171.0 |
| Case3 | −110.0 | −140.0 | −160.0 |
| Case4 | −98.0 | −152.0 | −173.0 |
| Case5 | −115.0 | −123.0 | −143.0 |
| Case6 | −101.0 | −152.0 | −178.0 |
| Case7 | −134.0 | −142.0 | −151.0 |
| Case8 | −123.0 | −165.0 | −177.0 |
| Case9 | −119.0 | −135.0 | −151.0 |
| Case10 | −132.0 | −154.0 | −175.0 |

**Table 4.**